\documentclass{ptephy_v2}

\preprintnumber{XXXX-XXXX} 
\usepackage{hyperref}

\newcommand{\cns}{Center for Nuclear Study, University of Tokyo, Hongo 7-3-1, Bunkyo, Tokyo 113-0033, Japan}
\newcommand{\kyoto}{Department of Physics, Kyoto University, Kitashirakawa, Oiwakecho, Sakyo, Kyoto 606-8502, Japan}
\newcommand{\riken}{RIKEN Nishina Center for Accelerator-Based Science, Hirosawa 2-1, Wako, Saitama 351-0198, Japan}
\newcommand{\rcnp}{Research Center for Nuclear Physics, The University of Osaka, 10-1 Mihogaoka, Ibaraki, Osaka 567-0047, Japan}
\newcommand{\kyushue}{Department of Advanced Energy Engineering Science, Kyushu University, Kasuga, Fukuoka 816-8580, Japan}
\newcommand{\rikkyo}{Department of Physics, Rikkyo University, Nishi-Ikebukuro, Tokyo 171-8501, Japan}
\newcommand{\ibs}{Institute for Rare Isotope Science, Institute for Basic Science, 1 Gukjegwahak-ro, Yuseong-gu, Daejeon 34000, Republic of Korea}
\newcommand{\kyushup}{Department of Physics, Kyushu University, Nishi, Fukuoka 819-0395, Japan}
\newcommand{\aizu}{Center for Mathematics and Physics, University of Aizu, Aizu-Wakamatsu, Fukushima 965-8580, Japan}
\newcommand{\gsi}{GSI Helmholtzzentrum f\"{u}r Schwerionenforschung, 64291 Darmstadt, Germany}

\newcommand\PPNP[1]{Prog.\ Part.\ Nucl.\ Phys.,\ \andvol{#1}}
\newcommand\NIMA[1]{Nucl.\ Instrum.\ Methods\ Phys.\ Res.\ Sect.\ A,\ \andvol{#1}}
\newcommand\NIMB[1]{Nucl.\ Instrum.\ Methods\ Phys.\ Res.\ Sect.\ B,\ \andvol{#1}}
\newcommand\PTEP[1]{Prog.\ Theor.\ Exp.\ Phys.,\ \andvol{#1}}

\begin{document}


\title{
Parity-transfer
$({}^{16}{\rm O},{}^{16}{\rm F}(0^-,{\rm g.s.}))$
reaction as a selective probe of isovector 
$0^-$ states in nuclei
}

\author{
M.~Dozono$^{1,2}$\thanks{Email: dozono.masanori.6v@kyoto-u.ac.jp}, 
M.~Ichimura$^3$\thanks{Deceased}, 
S.~Michimasa$^2$\thanks{present address: \riken}, 
M.~Takaki$^2$, 
M.~Kobayashi$^2$, 
M.~Matsushita$^2$, 
S.~Ota$^2$\thanks{present address: \rcnp}, 
H.~Tokieda$^2$, 
N.~Fukuda$^3$, 
N.~Inabe$^3$, 
S.~Kawase$^2$\thanks{present address: \kyushue}, 
K.~Kisamori$^3$, 
Y.~Kiyokawa$^2$, 
K.~Kobayashi$^4$, 
T.~Kubo$^3$, 
Y.~Kubota$^2$\thanks{present address: \riken}, 
C.~S.~Lee$^2$\thanks{present address: \ibs}, 
H.~Miya$^2$, 
A.~Ohkura$^5$, 
H.~Sagawa$^{3,6}$, 
S.~Sakaguchi$^5$, 
H.~Sakai$^3$, 
M.~Sasano$^3$, 
S.~Shimoura$^{2,3}$, 
Y.~Shindo$^5$, 
L.~Stuhl$^3$\thanks{present address: \gsi},
H.~Suzuki$^3$, 
H.~Tabata$^5$, 
H.~Takeda$^3$, 
T.~Wakasa$^5$, 
K.~Yako$^2$, 
Y.~Yanagisawa$^3$, 
J.~Yasuda$^5$, 
R.~Yokoyama$^2$, 
K.~Yoshida$^3$, 
J.~Zenihiro$^{1,3}$, 
and T.~Uesaka$^3$
}

\affil{
$^1${\kyoto}\\
$^2${\cns}\\
$^3${\riken}\\
$^4${\rikkyo}\\
$^5${\kyushup}\\
$^6${\aizu}\\
}






\begin{abstract}%
  We demonstrate that the parity-transfer
  $({}^{16}{\rm O},{}^{16}{\rm F}(0^-,{\rm g.s.}))$ reaction
  provides a selective probe of isovector $0^-$ excitations in nuclei.
  This reaction selectively populates unnatural-parity states
  through a $0^+ \to 0^-$ transition in the projectile.
  The excitation-energy spectrum of $^{12}\mathrm{B}$ was reconstructed 
  via the ${}^{12}{\rm C}({}^{16}{\rm O},{}^{16}{\rm F}(0^-,{\rm g.s.}))$ reaction at 247~MeV/u
  from the coincident detection of the ${}^{15}\mathrm{O}+p$ decay products of ${}^{16}\mathrm{F}$. 
  The known $0^{-}$ state at $E_x = 9.3~{\rm MeV}$ was clearly observed 
  with an enhanced forward cross section,
  confirming the selectivity of the reaction. 
  The structures observed at $E_x = 6.6 \pm 0.4$ and $14.8 \pm 0.3~{\rm MeV}$ 
  exhibit forward-peaked angular distributions 
  and are suggested to contain significant $0^-$ strength. 
  These results demonstrate that
  the parity-transfer reaction provides
  a powerful probe of $0^-$ excitations 
  and highlight its potential for systematic studies of spin-isospin modes,
  including pion-related dynamics, in nuclei. 
\end{abstract}

\subjectindex{D06, D13, D23}

\maketitle

\section{Introduction}
\label{sec:introduction}

The spin–isospin response of nuclei provides essential information on
the spin–isospin–dependent interaction in the nuclear medium~\cite{Osterfeld1992,Ichimura2006}. 
In particular,
this sector of the nuclear interaction is strongly influenced
by the pion-exchange interaction, 
reflecting the isovector ($T=1$) and
pseudoscalar nature ($J^{\pi}=0^{-}$) of the pion. 
Consequently, studies of spin–isospin excitations offer 
a sensitive probe of pion-related dynamics in nuclei. 

Among such excitations,
the isovector $0^{-}$ states are of particular interest
because they share the same quantum numbers as the pion.
Their properties have long been discussed in connection with
the possible softening of the pion mode in nuclei,
which has been suggested as a precursor phenomenon of
pion condensation in nuclear matter~\cite{Fayans1997,Migdal1978,Meyer-Ter-Vehn1981}. 
Experimental investigations of $0^{-}$ states, therefore, provide valuable insight into 
the behavior of the pion degree of freedom in finite nuclei.

Despite their importance,
experimental information on $0^{-}$ states remains limited
because their identification in reaction spectra is difficult.
The spin–dipole (SD) excitation, described by the operator $\sigma \tau r Y_1$,
consists of three components with $J^{\pi}=0^{-}$, $1^{-}$, and $2^{-}$.
Since these excitations share the same orbital angular momentum transfer ($L=1$),
the $0^{-}$ component is often obscured by stronger neighboring excitations,
particularly the $1^{-}$ and $2^{-}$ states~\cite{Osterfeld1992}.
Although polarization observables in reactions
such as $(p,n)$~\cite{Dozono2008,Wakasa2012}
and $(d,{}^{2}{\rm He})$~\cite{Okamura1995,Okamura2002,DeHuu2007} 
are known to be sensitive to the spin–parity of excited states and
have been used to disentangle these components, 
the extraction of the $0^{-}$ strength remains experimentally challenging.

\begin{figure} [t]
\begin{center}
  \includegraphics[width=12cm,clip]{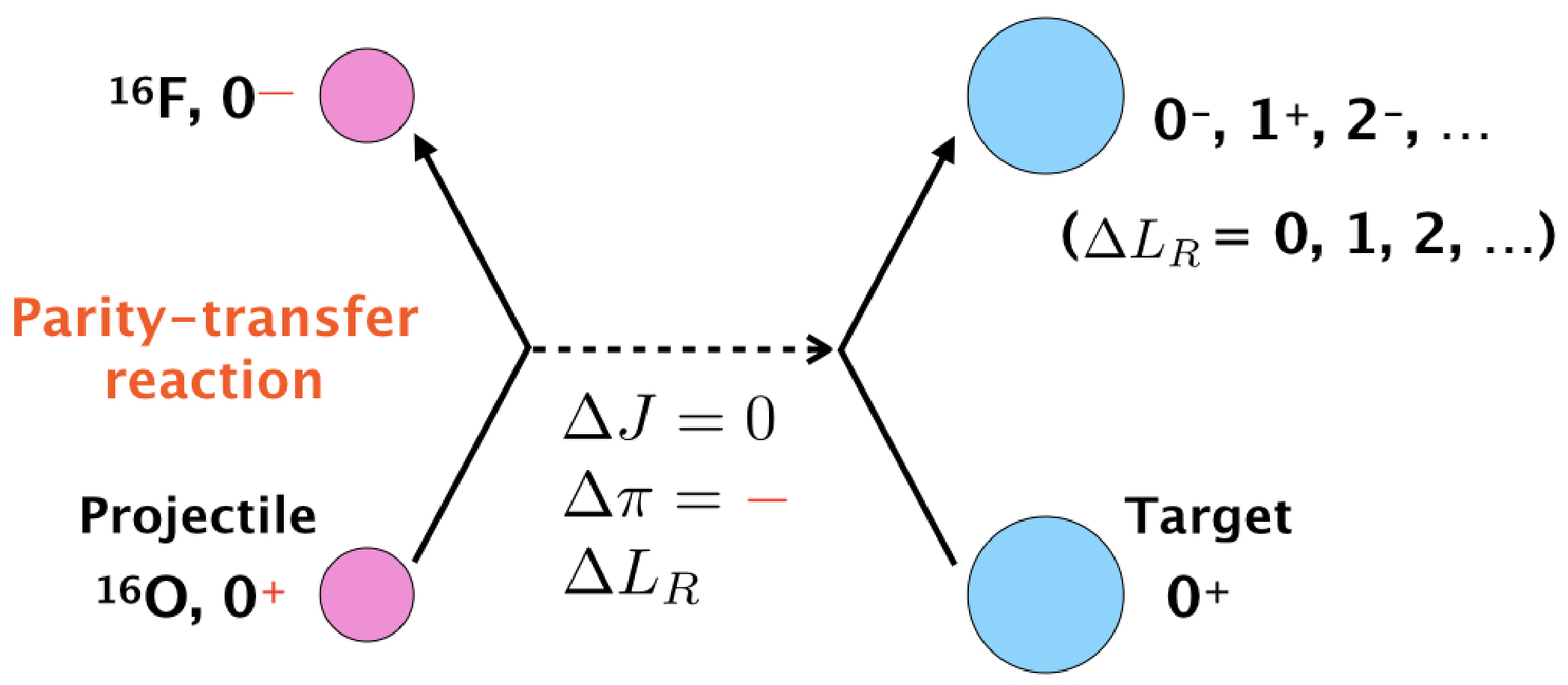}
\end{center}
\caption{
  Schematic illustration of the parity-transfer reaction on an even–even target nucleus. 
  The reaction exploits the $0^{+} \rightarrow 0^{-}$ transition in the projectile,
  effectively transferring internal parity to the target nucleus. 
  Because parity is conserved, 
  unnatural-parity states are selectively populated in the residual nucleus.
  In this reaction, the projectile spin is unchanged, 
  and the transferred orbital angular momentum $\Delta L_R$ 
  of the relative motion between the projectile and target 
  uniquely determines the spin–parity of the populated states,
  resulting in a one-to-one correspondence between
  $\Delta L_R = 0, 1, 2, \ldots$ and
  $J^\pi = 0^-, 1^+, 2^-, \ldots$, respectively.
  This property enables a clear identification of
  the $J^\pi$ components from the angular distributions
  and underlies the strong selectivity for the $0^-$ 
  component at forward angles, as discussed in the text.
}
\label{fig:parity_transfer_illustration}
\end{figure}

A promising approach to overcome this difficulty is
to utilize nuclear reactions that provide strong selectivity
for unnatural-parity states. 
In this study, 
we focus on target nuclei with $0^+$ ground states 
and investigate 
the parity-transfer $(^{16}\mathrm{O},\,^{16}\mathrm{F}(0^{-},\mathrm{g.s.}))$
reaction as such a probe.
As illustrated in Fig.~\ref{fig:parity_transfer_illustration}, 
this reaction exploits the $0^{+} \rightarrow 0^{-}$ transition in the projectile
and effectively transfers internal parity to the target nucleus. 
Because parity is conserved in the reaction, 
unnatural-parity states are selectively 
populated in the residual nucleus.

Moreover, 
the angular distributions depend sensitively on the spin–parity of the final states.
In the present reaction, the spin of the projectile is not changed,
and therefore the transferred orbital angular momentum $\Delta L_R$ 
of the relative motion between the projectile and the target 
uniquely determines the spin-parity of the populated states.
This one-to-one correspondence enables 
a clear identification of the $J^{\pi}$ components from the angular distributions. 
Among these, 
the present reaction exhibits 
a particularly strong selectivity
for the $0^-$ component at forward angles,
as demonstrated in this work. 
These features make the parity-transfer reaction a promising probe
for identifying and isolating $0^-$ excitations in nuclei. 

The parity-transfer 
$^{12}\mathrm{C}(^{16}\mathrm{O},\,^{16}\mathrm{F}(0^{-},\mathrm{g.s.}))$ reaction 
at 247~MeV/u was used to investigate the excitation energy spectrum of $^{12}\mathrm{B}$.
The $^{12}\mathrm{C}$ nucleus provides an ideal benchmark system
because a $0^{-}$ state at $E_x = 9.3~{\rm MeV}$ in $^{12}\mathrm{B}$
has been identified in previous polarization measurements~\cite{Okamura2002,DeHuu2007}.
By comparing the observed spectrum with theoretical calculations
and with results from previous experiments,
we investigate the effectiveness of the parity-transfer reaction as a probe for $0^-$ states in nuclei.

\section{Experiment and Data Analysis}
\label{sec:experiment}

The experiment was carried out at 
the RIKEN RI Beam Factory (RIBF)~\cite{Yano2007} 
using the SHARAQ spectrometer and 
a high-resolution beamline~\cite{Uesaka2012}. 
A primary ${}^{16}{\rm O}$ beam was 
accelerated to 247~${\rm MeV/{\rm nucleon}}$ 
and transported to the target position. 
The dispersion of the beamline was matched to that of the spectrometer 
to obtain an optimized excitation energy resolution~\cite{Michimasa2013,Dozono2016}.
The active ${}^{12}{\rm C}$ target was 
a $1~{\rm mm}$-thick plastic scintillation detector 
(equivalent ${}^{12}{\rm C}$ target thickness $= 103~{\rm mg/cm^2}$). 
This detector was horizontally segmented into 
16 plastic scintillators 
of area $5~{\rm mm^{\rm H}} \times 30~{\rm mm}^{\rm V}$.
From the hit pattern of the segments, 
we determined the horizontal position of the beam on the target. 
The beam intensity was indirectly monitored 
by a plastic counter 
installed after the target 
outside the acceptance region of the SHARAQ spectrometer.  
The beam intensity was typically maintained at approximately
$8 \times 10^{6}$~particles per second (pps), 
below the maximum intensity of $1 \times 10^{7}$~pps 
allowed by the radiation-safety regulations at the RIBF.

To measure the outgoing proton-unbound ${}^{16}{\rm F}$, 
the SHARAQ spectrometer was operated in 
``separated flow'' mode~\cite{Dozono2016}.
The experimental setup and analysis procedure
are detailed in Ref.~\cite{Dozono2016}. 
In this mode, 
the outgoing ${}^{15}{\rm O}+p$ pair produced 
in the ${}^{16}{\rm F}$ decay 
was separated by the first dipole magnet 
and detected at two focal planes of the SHARAQ. 
The ${}^{15}{\rm O}$ particle was detected with 
two low-pressure multi-wire drift chambers (LP-MWDCs)~\cite{Miya2013} 
and plastic scintillation counters at the final focal plane S2;
meanwhile, the proton was detected with two MWDCs 
and plastic scintillation counters at focal plane S1, 
which is located on the low-rigidity side downstream of 
the first dipole magnet.
The spectrometer was fixed at $0^{\circ}$, 
covering reaction angles up to $\theta_{\rm lab} \simeq 1^{\circ}$.

\begin{figure} [t]
\begin{center}
  \includegraphics[width=15cm,clip]{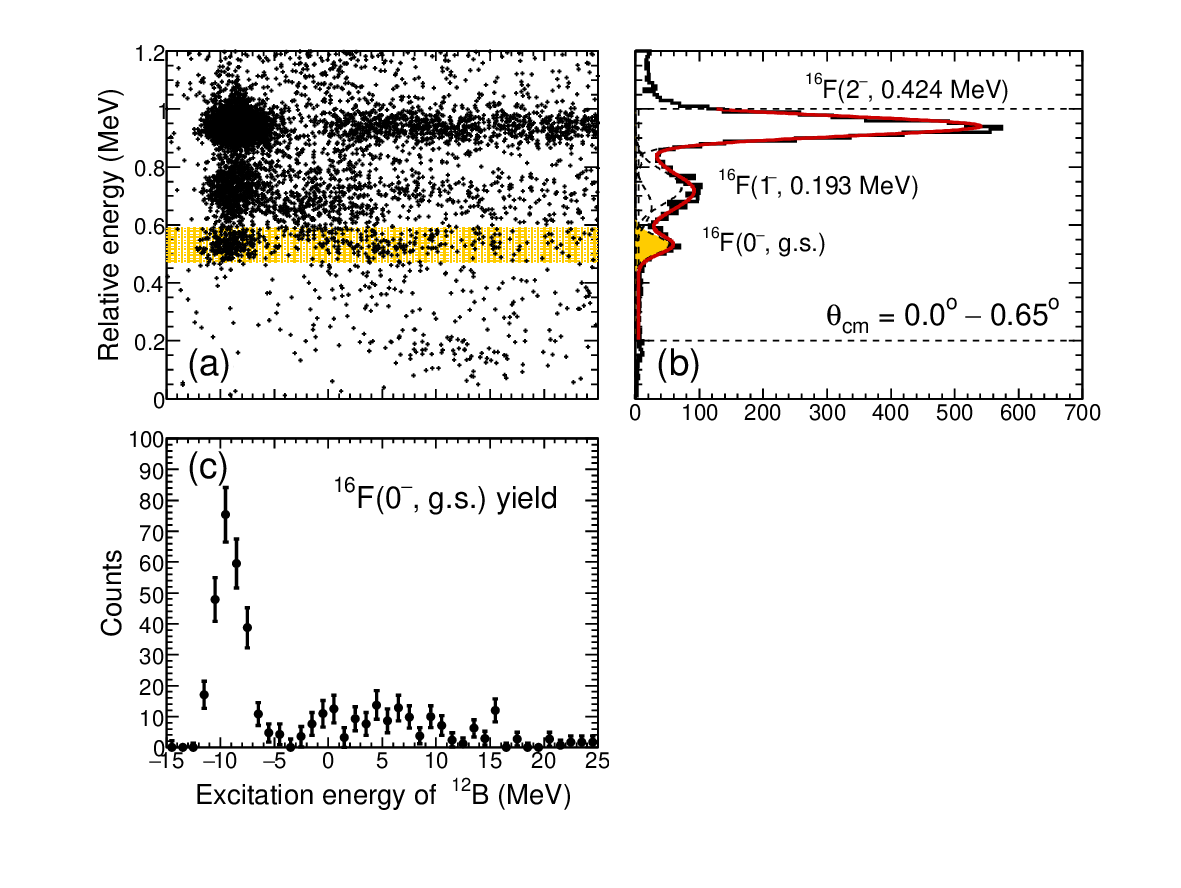}
\end{center}
\caption{
  (a) Correlation between the relative energy $E_{\mathrm{rel}}$
  of the ${}^{15}\mathrm{O}+p$ system 
  and the excitation energy $E_x$ of ${}^{12}\mathrm{B}$ 
  for $\theta_{\mathrm{cm}} = 0.0^{\circ}–0.65^{\circ}$,
  shown as a representative example. 
  (b) Projection of (a) onto $E_{\mathrm{rel}}$. 
  The spectrum is fitted with contributions from
  the $0^-$ g.s. of ${}^{16}\mathrm{F}$, 
  the $1^-$ state at 0.193~MeV,
  and the $2^-$ state at 0.424~MeV. 
  A Gaussian component at $E_{\mathrm{rel}} = 0.664~{\rm MeV}$
  and a constant background are 
  also included to account for additional contributions, 
  possibly associated with decays from higher-lying states in ${}^{16}\mathrm{F}$. 
  (c) Excitation-energy spectrum of ${}^{12}\mathrm{B}$
  for the ${}^{16}\mathrm{F}(0^-, {\rm g.s.})$ channel, 
  obtained by fitting the $E_{\mathrm{rel}}$ distribution in each $E_x$ bin.
}
\label{fig:ex_vs_erel}
\end{figure}

The momentum vectors of the outgoing proton and ${}^{15}\mathrm{O}$ 
at the target were reconstructed from the positions and angles 
measured at the S1 and S2 focal planes using a ray-tracing method. 
From the reconstructed momentum vectors,
the relative energy $E_{\mathrm{rel}}$ of the ${}^{15}\mathrm{O}+p$ system
was determined using the invariant-mass method,
while the excitation energy $E_x$ of ${}^{12}\mathrm{B}$
was obtained from the missing-mass method. 
The reaction angle $\theta_{\mathrm{cm}}$ was also determined event by event. 

To isolate the ${}^{16}\mathrm{F}(0^-, {\rm g.s.})$ channel,
the following analysis was performed for each angular bin. 
As a representative example, 
the results for the forward-angle region
$(\theta_{\mathrm{cm}} = 0.0^{\circ}–0.65^{\circ})$ are shown in Fig.~\ref{fig:ex_vs_erel}. 
Figure~\ref{fig:ex_vs_erel}(a) 
shows the correlation between 
$E_{\mathrm{rel}}$ of the ${}^{15}\mathrm{O}+p$ system
and the excitation energy $E_x$ of ${}^{12}\mathrm{B}$. 
Different loci corresponding to 
the decay of ${}^{16}\mathrm{F}$ are observed,
indicating a clear kinematic separation of the ${}^{16}\mathrm{F}$ states. 

The $E_{\mathrm{rel}}$ spectrum for this angular range 
is shown in Fig.~\ref{fig:ex_vs_erel}(b).
Owing to the energy resolution of 80~keV 
in full width at half maximum (FWHM), 
the spectrum exhibits well-separated peaks corresponding to 
the known low-lying states in ${}^{16}\mathrm{F}$, 
namely the $0^-$ ground state ($E_{\rm rel} = 0.535~{\rm MeV}$),
the $1^-$ state at 0.193~MeV ($E_{\rm rel} = 0.728~{\rm MeV}$),
and the $2^-$ state at 0.424~MeV ($E_{\rm rel} = 0.959~{\rm MeV}$). 
The spectrum was fitted with these three components,
where the intrinsic widths were taken from Ref.~\cite{HFujita2009}, 
and folded with the experimental energy resolution. 
In addition to these components,
a Gaussian peak at $E_{\mathrm{rel}} = 0.664~{\rm MeV}$
and a constant background term were included 
to account for residual contributions.
These components may originate from 
the decay of higher-lying states in ${}^{16}\mathrm{F}$, 
leading to ${}^{15}\mathrm{O}^* + p$ final states. 

The excitation-energy spectrum of ${}^{12}\mathrm{B}$ 
corresponding to the ${}^{16}\mathrm{F}(0^-, {\rm g.s.})$ channel
was extracted by fitting the $E_{\mathrm{rel}}$ spectrum in each $E_x$ bin.
The resulting spectrum is shown in Fig.~\ref{fig:ex_vs_erel}(c).
The excitation-energy resolution was $2.6~{\rm MeV}$ in FWHM. 
A peak is also observed at $E_x \sim -10~{\rm MeV}$. 
This structure originates from reactions with hydrogen contained in the target material 
and does not correspond to states in ${}^{12}\mathrm{B}$.

The same fitting procedure was applied to all angular bins. 
After correction for the detection efficiency 
of the ${}^{15}{\rm O}+p$ coincidence events, 
estimated by Monte Carlo simulations,
the double-differential cross sections for 
the $({}^{16}{\rm O},{}^{16}{\rm F}(0^-,{\rm g.s.}))$ reaction
were extracted. 
The detection efficiency (19\%) was mainly limited by
the angular acceptance range of the protons. 

\section{Results}
\label{sec:results}

\begin{figure} [t]
\begin{center}
    \includegraphics[width=12cm,clip]{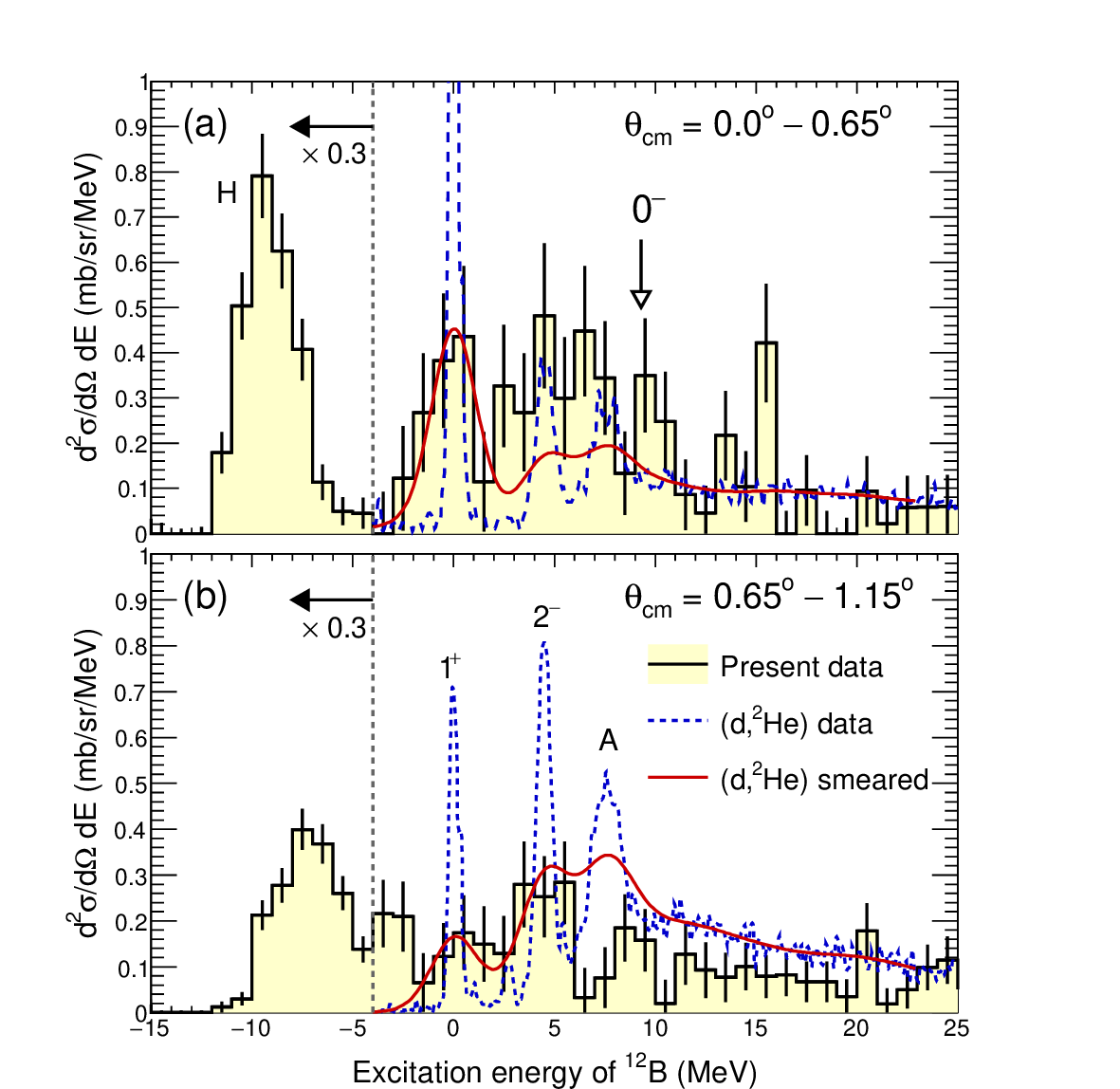}
\caption{
  Double differential cross sections of 
  the ${}^{12}{\rm C}({}^{16}{\rm O},{}^{16}{\rm F}(0^-,{\rm g.s.}))$
  reaction at 
  (a) $\theta_{\rm cm}=0^{\circ}$--$0.65^{\circ}$ and 
  (b) $\theta_{\rm cm}=0.65^{\circ}$--$1.15^{\circ}$. 
  Dashed curves are the experimental data 
  of the ${}^{12}{\rm C}(d,{}^{2}{\rm He})$ reaction 
  at $270~{\rm MeV}$~\cite{Okamura1995}. 
  The $(d,{}^{2}{\rm He})$ spectra (solid curves)  
  are smeared out to match the energy resolution of our data. 
  See text for details. 
  }
\label{fig:ex_spectra}
\end{center}
\end{figure}

Figure~\ref{fig:ex_spectra}(a) and (b) show 
the double differential cross sections 
of the ${}^{12}{\rm C}({}^{16}{\rm O},{}^{16}{\rm F}(0^-,{\rm g.s.}))$ reaction 
at $\theta_{\rm cm}=0^{\circ}-0.65^{\circ}$ and $0.65^{\circ}-1.15^{\circ}$, 
respectively. 
The hydrogen-induced peak at $E_x \sim -10~{\rm MeV}$ is 
clearly separated from the ${}^{12}\mathrm{B}$ excitation region at forward angles. 
Owing to the kinematic shift with increasing scattering angle, 
the hydrogen-related events move toward higher excitation energies 
and gradually approach the ${}^{12}\mathrm{B}$ excitation region.  

To demonstrate the selectivity of the parity-transfer reaction,
the present results were overlaid on 
the previous data of 
the ${}^{12}{\rm C}(d,{}^{2}{\rm He})$ reaction 
at an incident energy of $270~{\rm MeV}$~\cite{Okamura1995}. 
The spectra at 
$\theta_{\rm cm}=0^{\circ}-1^{\circ}$ and $6^{\circ}-8^{\circ}$
are presented as the dashed curves in Fig.~\ref{fig:ex_spectra}(a) and (b), 
respectively.
In both cases, the momentum transfers were comparable 
($q \sim 0.3$ and $0.5~{\rm fm}^{-1}$
for Fig.~\ref{fig:ex_spectra}(a) and (b), respectively). 
The $(d,{}^{2}{\rm He})$ cross sections, 
plotted as solid curves, 
were smeared to 
match our energy resolution. 
The $(d,{}^{2}{\rm He})$ spectra were scaled 
to match the $({}^{16}{\rm O},{}^{16}{\rm F}(0^-,{\rm g.s.}))$ 
cross sections of $1^+$ g.s..

Excitation of the $1^+$ g.s. and 
the $2^-$ state at $E_x=4.5~{\rm MeV}$ 
is observed in both reactions, 
but the structures at $E_x \gtrsim 6~{\rm MeV}$ 
are markedly different between the two reactions. 
The peak at $E_x = 7.5~{\rm MeV}$ (labeled ``A'' in Fig.~\ref{fig:ex_spectra}(b)) is
prominent in the $(d,{}^{2}{\rm He})$ data,
but is barely observable in the $({}^{16}{\rm O},{}^{16}{\rm F}(0^-,{\rm g.s.}))$ data.
This difference is discussed in detail in Sec~\ref{sec:discussion3}. 

Another striking difference is seen at $E_x \sim 9~{\rm MeV}$ 
in Fig.~\ref{fig:ex_spectra}(a); 
the clear enhancement in the 
$({}^{16}{\rm O},{}^{16}{\rm F}(0^-,{\rm g.s.}))$ data
vanishes in the $(d,{}^{2}{\rm He})$ data. 
This enhancement is attributed to the known $0^-$ state at $E_{x}=9.3~{\rm MeV}$,  
which was found only with the help of tensor analyzing powers 
of the $(d,{}^{2}{\rm He})$ reaction~\cite{Okamura2002}, 
indicating the selectivity of the present reaction for $0^-$ states. 
A similar enhancement at $E_x \sim 15~{\rm MeV}$
is a potential candidate for a new $0^-$ state. 

\section{Discussions}
\label{sec:discussions}

In this section,
we discuss the observed features of the excitation-energy spectrum
and angular distributions
in terms of the reaction selectivity and the underlying spin–parity structure.

\subsection{Reaction selectivity and DWBA+SM interpretation}
\label{sec:discussion1}

\begin{figure} [t]
\begin{center}
      \includegraphics[width=12cm,clip]{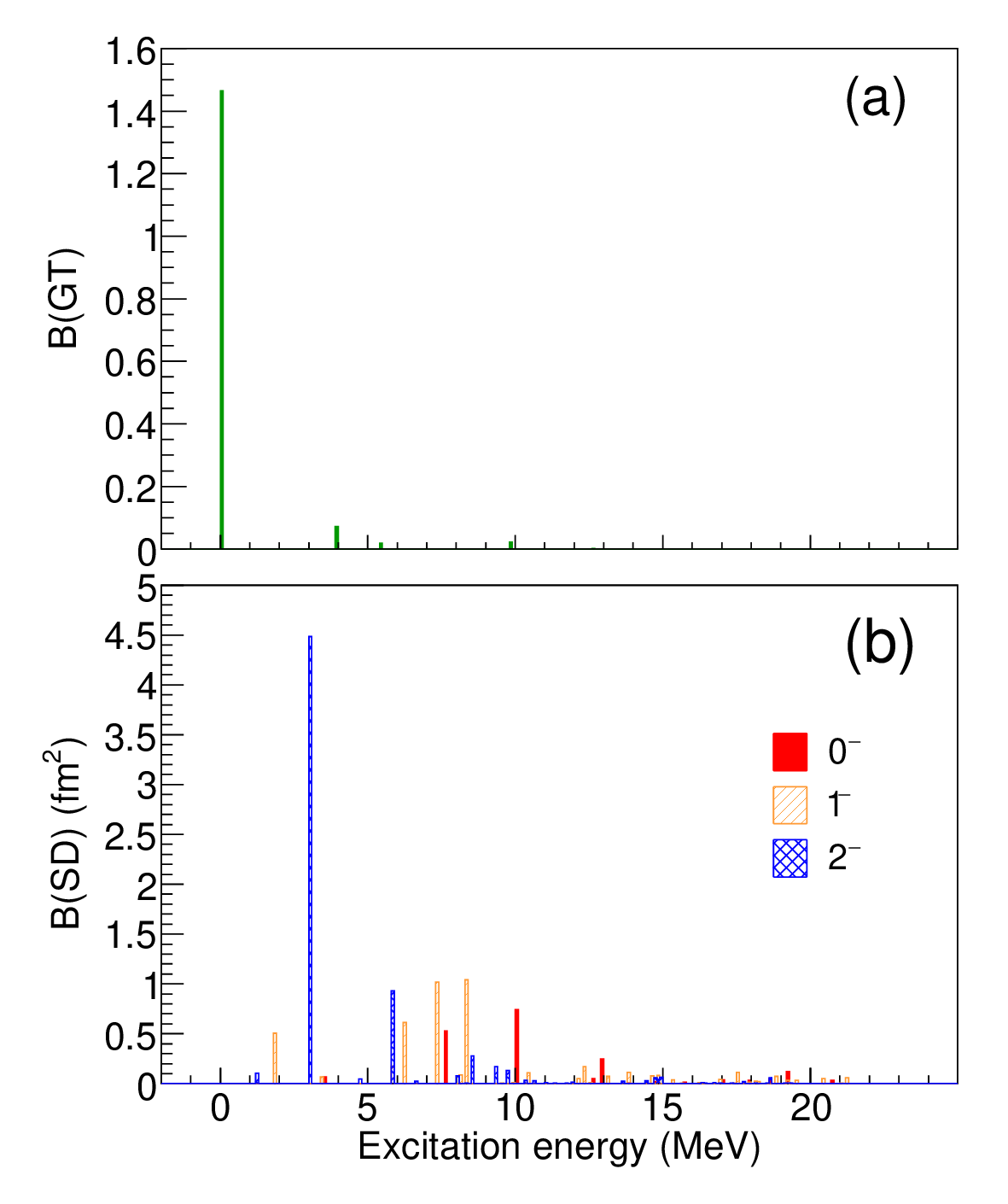} 
\caption{
  SM transition strengths calculated
  with the WBT interaction~\cite{Warburton1992}. 
  (a) Gamow–Teller ($1^+$) strength distribution B(GT). 
  (b) SD strength distribution B(SD) 
  for the $0^-$ (solid),
  $1^-$ (hatched), and $2^-$ (cross-hatched) components
  as a function of excitation energy in ${}^{12}\mathrm{B}$.
  }
\label{fig:WBT}
\end{center}
\end{figure}

Figure~\ref{fig:WBT} shows 
the Gamow–Teller ($1^+$) and SD ($0^-$, $1^-$, and $2^-$) strength 
distributions calculated with the shell model (SM)
using the WBT interaction~\cite{Warburton1992}. 
The Gamow–Teller strength is predominantly concentrated in the g.s. transition,
whereas the SD strength is distributed over a 
wide excitation-energy region, 
particularly above $E_x \sim 4~{\rm MeV}$. 
The overall features of these distributions are consistent with 
those observed in previous charge-exchange reactions 
such as $(n,p)$ and $(d,{}^{2}{\rm He})$~\cite{Okamura1995,Okamura2002,DeHuu2007,Olsson1993,Yang1993}.

\begin{figure} [t]
\begin{center}
    \includegraphics[width=12cm,clip]{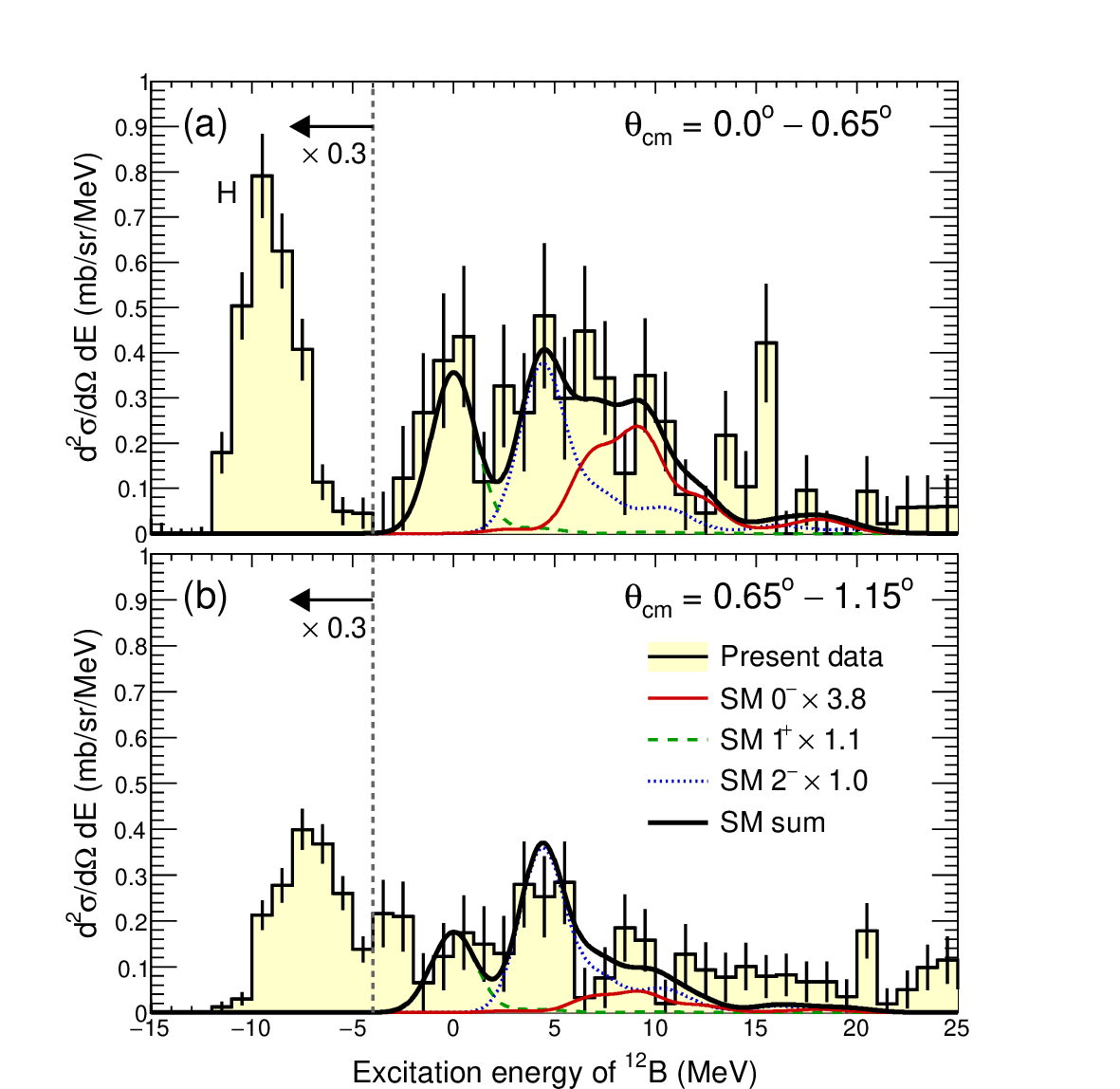}
\caption{
  Double differential cross sections of
  the $^{12}\mathrm{C}(^{16}\mathrm{O},\,^{16}\mathrm{F}(0^{-},\mathrm{g.s.}))$
  reaction (histograms) compared with
  DWBA+SM calculations (thick solid lines).
  Contributions from different spin–parity components are also shown:
  $0^{-}$ (solid), $1^{+}$ (dashed), and $2^{-}$ (dotted).
  See text for details of the normalization procedure. 
  }
\label{fig:sm_spectra}
\end{center}
\end{figure}

To examine whether the observed excitation-energy spectrum reflects 
the expected selectivity of the parity-transfer reaction, 
we compared the data with 
distorted-wave Born approximation (DWBA) calculations 
using the SM transition strengths shown in Fig.~\ref{fig:WBT} as input. 
In the following, this combined framework is referred to as DWBA+SM. 
The DWBA calculations were performed for the $0^-$, $1^+$, and $2^-$ components 
using the reaction code {\sc fold/dwhi}~\cite{FOLDDWHI},
with the Franey-Love nucleon-nucleon interaction at 270~MeV~\cite{Franey1985}.
The One-body transition densities for both the projectile and target systems
were taken from the SM calculations. 
The optical-model potentials used to generate the distorted waves 
were taken from the global optical potential of Ref.~\cite{Furumoto2012}, 
which was constructed within a double-folding model using the CEG07b G-matrix interaction.

The calculated double-differential cross sections
were folded with the experimental angular resolution
(3~mrad in FWHM) and
integrated over the corresponding angular ranges to obtain the spectra. 
The resulting spectra were then folded with
the experimental energy resolution (2.6~MeV in FWHM). 

Since the absolute cross sections are subject to
uncertainties in the DWBA+SM framework,
including both the reaction mechanism and the structure input, 
normalization factors were introduced separately for each spin–parity component 
by reproducing the known states at
$E_x = 0.0~{\rm MeV} (1^+)$, $4.5~{\rm MeV} (2^-)$, and $9.3~{\rm MeV} (0^-)$.
The same normalization factors were applied to all angular bins. 
The resulting normalization factors were 1.1, 1.0, and 3.8
for the $1^+$, $2^-$, and $0^-$ components, respectively.
In addition, small energy shifts were introduced 
to align the reference states used for the normalization.
The required shifts were
$+1.35~{\rm MeV}$ for the $2^-$ component and $-0.8~{\rm MeV}$ for the $0^-$ component
relative to the shift applied to the $1^+$ component.

The large normalization required for the $0^-$ component may 
reflect limitations of the present DWBA description of the reaction mechanism.
In particular, the exchange term is treated approximately in the present framework,
and the tensor-force exchange contribution is neglected~\cite{FOLDDWHI}, 
which may affect the calculated cross sections. 
Uncertainties in the structure description of the transitions 
may also affect the normalization factors. 
A quantitative understanding of the normalization factors,
especially for the $0^-$ component, is beyond the scope of the present work and
remains as a subject for future study. 
The present comparison should therefore be regarded as qualitative. 

Figure~\ref{fig:sm_spectra} compares the experimental spectra
with those predicted by the DWBA+SM calculation. 
The overall spectral features are reasonably well reproduced by the calculations for both angular ranges. 
This agreement suggests that the observed excitation-energy spectrum follows 
the pattern expected for the parity-transfer reaction,
which selectively populates unnatural-parity states.
In the experimental spectrum,
the cross section at forward angles increases above $E_x \sim 6~{\rm MeV}$. 
The DWBA+SM calculation predicts that
this excitation-energy region is dominated by the $0^-$ component,
suggesting that the observed enhancement is largely associated with $0^-$ contributions. 

To further clarify the spin–parity nature of individual structures
suggested by the DWBA+SM interpretation, 
a quantitative decomposition of the spectrum
is presented in the following subsection. 

\subsection{Spin-parity analysis of the observed structures}
\label{sec:discussion2}

\begin{figure} [t]
\begin{center}
      \includegraphics[width=12cm,clip]{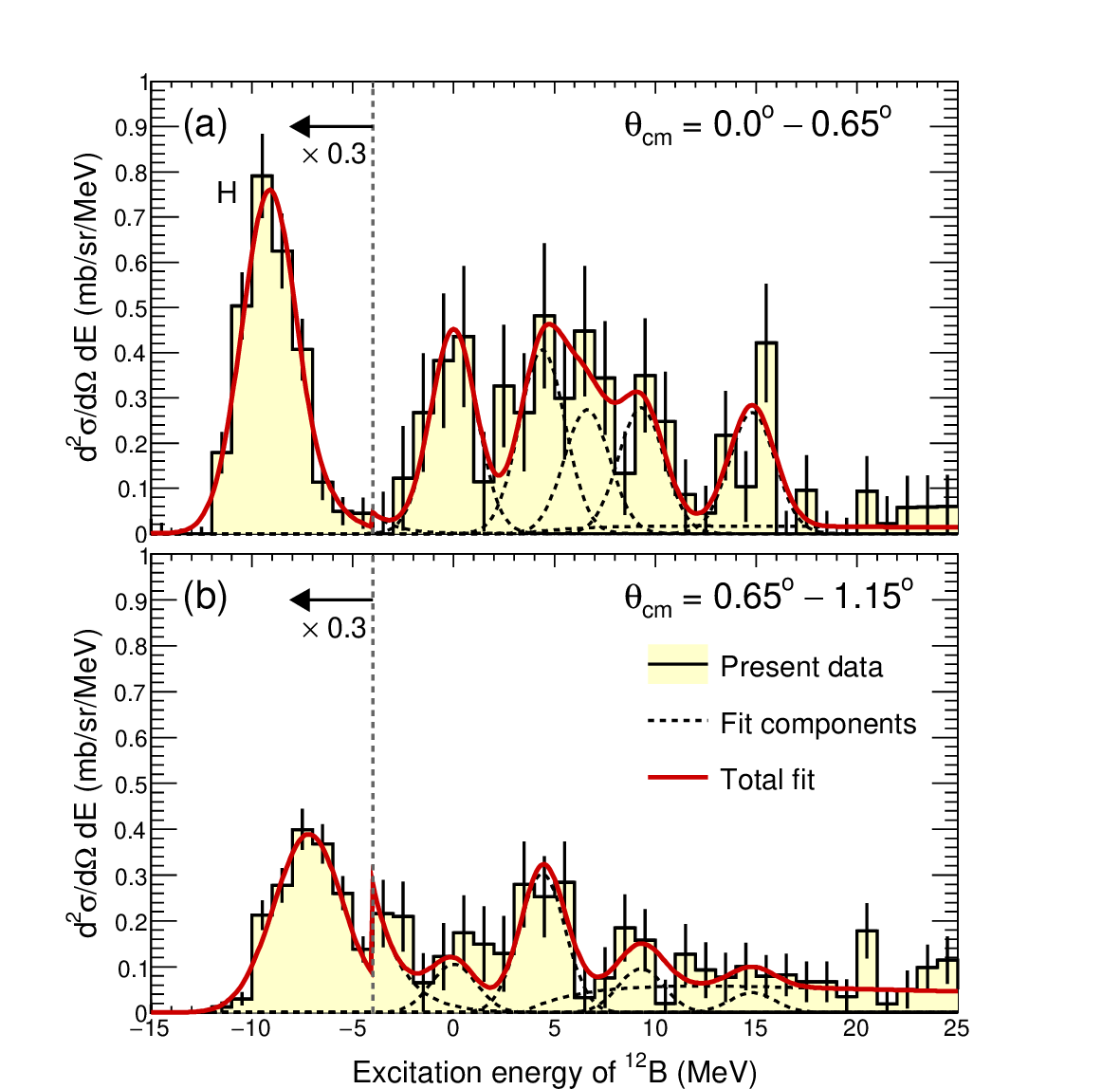} 
      \caption{
        Peak-fitting results of the excitation-energy spectra for the
        ${}^{12}\mathrm{C}({}^{16}\mathrm{O},{}^{16}\mathrm{F}(0^-, {\rm g.s.}))$ reaction. 
        (a) $\theta_{\mathrm{cm}} = 0^{\circ}–0.65^{\circ}$ and
        (b) $\theta_{\mathrm{cm}} = 0.65^{\circ}–1.15^{\circ}$. 
        The solid curves show the total fit, while the dashed curves indicate 
        individual fit components. 
  }
\label{fig:fit}
\end{center}
\end{figure}

To investigate the spin–parity of individual structures in more detail,
the cross sections of individual components were extracted by fitting
the excitation-energy spectra with a set of Gaussian peaks.
The continuum background arising from quasi-free scattering events 
was estimated by the formula in Ref.~\cite{Erell1986}, 
and the parameters were taken from Ref.~\cite{Okamura2002}. 
The ${}^{1}{\rm H}({}^{16}{\rm O},{}^{16}{\rm F}(0^-,{\rm g.s.}))$ 
background at $E_{x} < 0~{\rm MeV}$ 
was described as a Gaussian peak with an exponential tail. 
Three known states with unnatural parity, 
namely the $1^+$ g.s., 
the $2^-$ state at $E_x = 4.5~{\rm MeV}$, 
and the $0^-$ state at $E_x = 9.3~{\rm MeV}$, 
were included as discrete components in the fit. 
In addition, two structures were observed
around $E_x \sim 7$ and $15~{\rm MeV}$,
and corresponding Gaussian components were introduced
to describe these features. 
The fitted peak positions were determined to be
$E_x = 6.6(4)~{\rm MeV}$ and $14.8(3)~{\rm MeV}$. 
For the peak widths, 
only the experimental energy resolution of 2.6 MeV was considered,
assuming that the intrinsic widths of the states are small
compared to the resolution. 
The peak positions and widths were fixed in the fitting procedure. 
Figure~\ref{fig:fit} shows 
the peak-fitting results of the spectra 
at $\theta_{\rm cm}=0^{\circ}-0.65^{\circ}$ (a)
and $\theta_{\rm cm}=0.65^{\circ}-1.15^{\circ}$ (b).

\begin{figure} [t]
\begin{center}
      \includegraphics[width=12cm,clip]{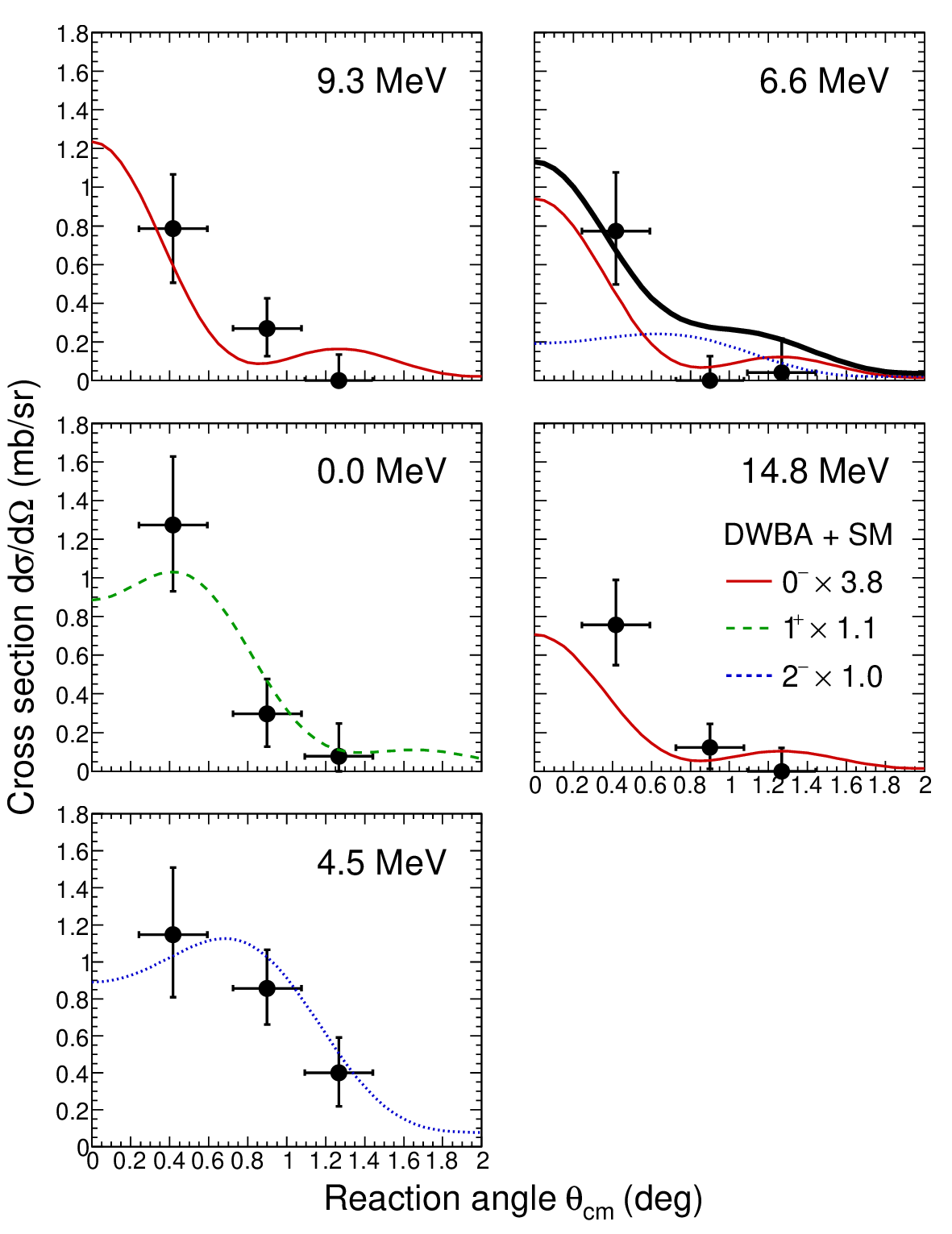} 
      \caption{
        Measured angular distributions of the differential cross sections
        for the indicated excitation energies in ${}^{12}\mathrm{B}$. 
        The solid, dashed, and dotted curves represent the DWBA calculations
        for the $0^-$, $1^+$, and $2^-$ components, respectively, 
        while the thick solid curve shows their sum. 
        The normalization factors of the DWBA cross sections
        are the same as those used in the DWBA+SM spectral comparison. 
        The SM states used in the DWBA calculations are: 
        $0^-_4$ (9.3~MeV), $1^+_1$ (g.s.), $2^-_2$ (4.5~MeV), 
        a mixture of $0^-_2$ and $2^-_4$ (6.6~MeV), 
        and the summed $0^-_{5-20}$ states (14.8~MeV). 
  }
\label{fig:angular_distributions}
\end{center}
\end{figure}

The angular distributions of the extracted cross section 
are shown in Fig.~\ref{fig:angular_distributions}. 
The $0^-$ state at $E_x = 9.3~{\rm MeV}$ 
shows a forward-peaking angular distribution, 
which enhances the signal-to-noise ratio of this state 
at the most forward reaction angle. 
The cross-section ratio of the $0^-$ state to the $1^+$ g.s.
reaches about 0.6 in the present reaction, 
whereas in the $(d,{}^{2}{\rm He})$ reaction 
it is much smaller than 0.1 
(e.g., see Fig.~1 in Ref.~\cite{Okamura2002}).
This comparison indicates the high efficiency of 
the parity-transfer reaction for investigating $0^-$ states. 

The DWBA results for 
the $0^-$, $1^+$, and $2^-$ states are also shown 
in Fig.~\ref{fig:angular_distributions}.
The same normalization factors as those determined
in the DWBA+SM spectral comparison were used. 
The calculations predict characteristic  
spin–parity–dependent oscillatory patterns in the angular distributions. 
In particular, the $0^-$ state exhibits a strong forward peak,
whereas the $1^+$ and $2^-$ states show their first maxima at finite angles. 
These patterns reproduce well 
the experimental data of the known states, 
the $0^-$ state at $E_x = 9.3~{\rm MeV}$, 
the $1^+$ g.s. and 
the $2^-$ state at $E_x = 4.5~{\rm MeV}$. 
Thus, the characteristic oscillatory pattern of the angular distributions 
provides information on the spin-parity of the excited states.

Using these characteristic features, 
we investigate the spin-parity structures at $6.6~{\rm MeV}$. 
According to the SM calculations shown in Fig.~\ref{fig:WBT}, 
this energy region is expected to 
contain both $0^-$ and $2^-$ transition strengths, 
with comparable contributions from the $0^-_2$ and $2^-_4$ states. 
The observed angular distribution exhibits forward-peaked behavior, 
which cannot be reproduced by a pure $2^-$ component,
particularly at forward angles. 
A better agreement is obtained when a $0^-$ contribution is included.
An illustrative comparison based on such a mixture
is shown in Fig.~\ref{fig:angular_distributions}. 

Additional fits were performed to examine the stability of this structure.
The significance of the 6.6-MeV component was found to 
depend on the fitting conditions,
reflecting the influence of nearby states
such as the $2^-$ state at $E_x = 4.5~{\rm MeV}$
and the $0^-$ state at $E_x = 9.3~{\rm MeV}$. 
Therefore, while the present data suggest that
the 6.6-MeV structure may contain a non-negligible $0^-$ component,
its existence and detailed composition remain tentative.

A similar analysis was performed for the structure observed at $E_x = 14.8~{\rm MeV}$. 
The angular distribution also shows a forward-peaked behavior, 
although the statistical uncertainty is relatively large. 
To examine the possible contribution of $0^-$ excitations in this energy region,
the experimental angular distribution was compared with DWBA calculations
obtained by summing the contributions of higher-lying $0^-$ states.

An illustrative comparison based on such a summed $0^-$ contribution
is shown in Fig.~\ref{fig:angular_distributions},
where the forward-peaked tendency of the data is reasonably reproduced. 
Although the present statistics do not allow a unique spin–parity assignment, 
the observed angular distribution suggests that 
the enhancement around $E_x =  14.8~{\rm MeV}$ 
is consistent with the presence of significant $0^-$ strength. 

The angular distributions of the structures at $E_x = 6.6$ and $14.8$ MeV
can also be reproduced, within the experimental uncertainties,
by DWBA curves for the $1^+$ component. 
However, this assignment appears unlikely.
SM calculations predict that
the $1^+$ strength is predominantly concentrated
in the g.s. transition
and show negligible strength in this excitation-energy region,
as clearly shown in Fig.~\ref{fig:WBT}.
In addition, previous charge-exchange experiments
do not exhibit pronounced structures in this region
characteristic of strong $1^+$ transitions. 
Taken together, these considerations suggest that
the forward-peaked structures observed at $E_x = 6.6$ and $14.8$ MeV
are more naturally interpreted as arising from $0^-$ strength, 
although a unique assignment cannot be established with the present statistics.

\subsection{Nature of the 7.5-MeV structure}
\label{sec:discussion3}

While the structures discussed above can be understood 
in terms of the spin–parity selectivity of the parity-transfer reaction, 
the present data also offer new insight into
the long-standing controversy on
the spin–parity of the bump structure at $E_x = 7.5$ MeV
(labeled “A” in Fig.~\ref{fig:ex_spectra}(b)).
The Uppsala and Los Alamos groups, 
who studied the cross section of 
the ${}^{12}{\rm C}(n,p)$ reaction, 
attributed this bump mainly to $J^{\pi} = 1^-$~\cite{Olsson1993,Yang1993}. 
This interpretation was based on the SM prediction that 
the transition strength around $E_x \sim 7~{\rm MeV}$ is 
dominated by the $1^-$ component (see Fig.~\ref{fig:WBT}), 
together with the good agreement between the measured angular distributions and 
theoretical calculations based on these strengths. 
The result was supported by
angular-distribution measurements
of the decay neutrons from residual $^{12}{\rm B}$ 
produced by the ${}^{12}{\rm C}(d,{}^{2}{\rm He})$
reaction, performed at RCNP~\cite{Inomata1998}. 
However, tensor analyzing powers of 
the ${}^{12}{\rm C}(\vec{d},{}^{2}{\rm He})$ 
reaction measured at RIKEN~\cite{Okamura1995,Okamura2002} 
suggested a main component of $J^{\pi}=2^-$. 
New polarization data from the high-resolution 
${}^{12}{\rm C}(\vec{d},{}^{2}{\rm He})$
reaction experiment at KVI~\cite{DeHuu2007} 
have refined the picture. 
The bump appears to comprise two components:
a low-energy part with $J^{\pi}=2^-$ and
a high-energy part with $J^{\pi}=1^-$. 
The inconsistency among the spin-parity assignments 
remains unsolved. 

Our data showed no apparent structure at $E_x=7.5~{\rm MeV}$, 
indicating that the structures seen in 
the $(n,p)$ and $(d,{}^{2}{\rm He})$ reactions 
are dominated by $1^-$ states.
This is expected 
because the parity-transfer reaction 
does not excite the natural parity state. 
Furthermore,
the possible presence of $0^-$ strength 
in the vicinity of $E_x \sim 6 - 7~{\rm MeV}$ 
suggests that a small $0^-$ admixture may be present
in the 7.5-MeV structure.
In the $(d,{}^{2}{\rm He})$ reaction,
the tensor analyzing powers of the $0^-$ and $1^-$ transitions
have opposite signs.
Therefore, a mixture of these components
can lead to a strong cancellation,
resulting in tensor analyzing powers close to zero.
Such a cancellation may lead to
an ambiguous spin-parity assignment
when tensor analyzing powers are used as the primary observables. 

To examine this effect quantitatively, 
reaction calculations were performed 
using the adiabatic coupled-channel Born approximation
(ACCBA) code~\cite{Okamura1999}. 
The parameters were taken from Ref.~\cite{Okamura1999}, 
except for the transition strengths,
which were obtained from SM calculations using the WBT interaction. 
The calculations were compared with the experimental data 
for the $^{12}{\rm C}(d,{}^{2}{\rm He})$ reaction at 270~MeV 
measured at RIKEN~\cite{Okamura1995,Okamura2002}.

\begin{figure} [t]
\begin{center}
      \includegraphics[width=7cm,clip]{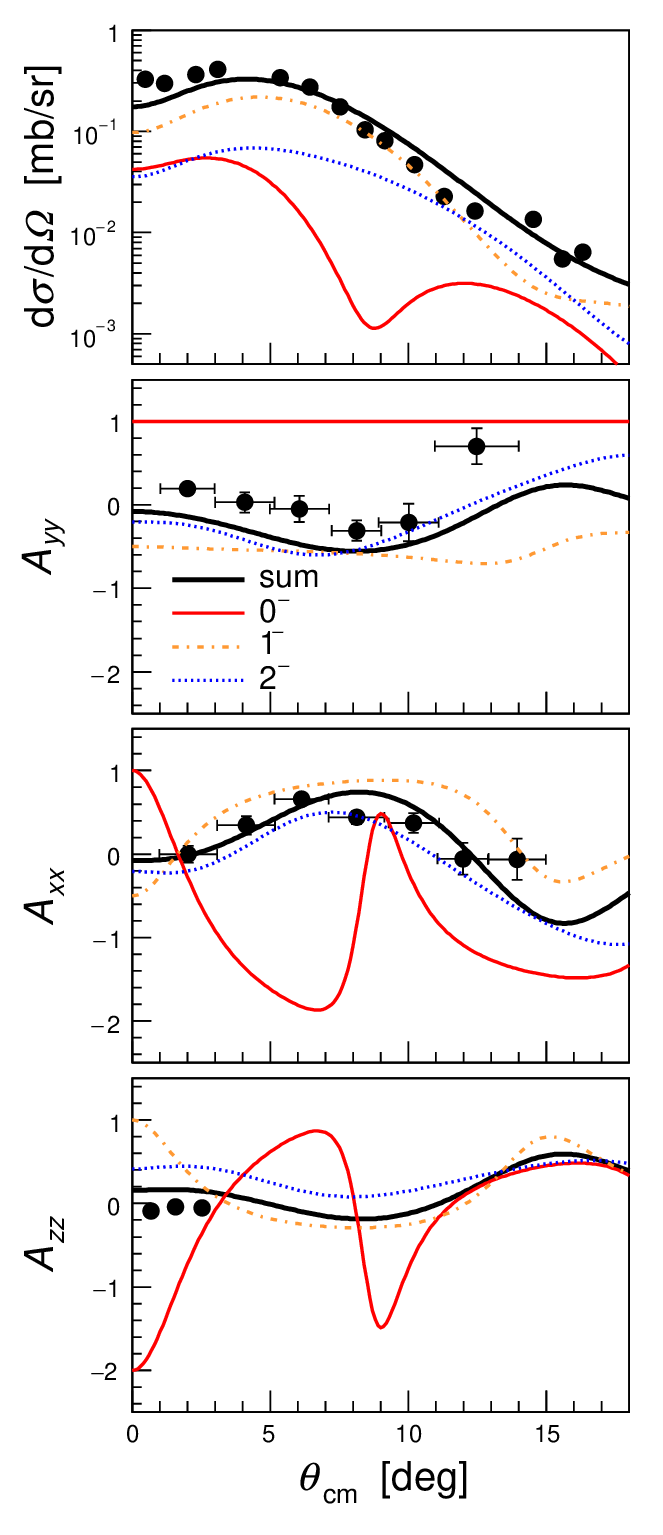} 
\caption{
  Cross section and tensor analyzing powers
  for the $^{12}{\rm C}(d,{}^{2}{\rm He})$ reaction
  populating the $E_x = 7.5~{\rm MeV}$ component in ${}^{12}{\rm B}$.
  The experimental data are taken from
  the RIKEN measurement at
  $E_d = 270~{\rm MeV}$~\cite{Okamura1995,Okamura2002}.
  The curves are calculated assuming contributions
  from representative SM states in this energy region
  (e.g., $0^-_2$, $1^-_{4-7}$ and $2^-_4$),
  and are shown for the $0^-$, $1^-$, and $2^-$ components
  and their sum.
  The normalization factors for the $0^-$ and $2^-$ components
  were determined to reproduce
  the known states at $E_x = 9.3~{\rm MeV}$ and $E_x = 4.5~{\rm MeV}$,
  respectively,
  while that of the $1^-$ component was adjusted
  to reproduce the bump structure around $E_x = 7.5~{\rm MeV}$. 
  }
\label{fig:accba}
\end{center}
\end{figure}

As an illustrative example, 
the cross sections and tensor analyzing powers
were calculated assuming contributions
from representative $0^-$, $1^-$ and $2^-$ states 
predicted by SM calculations 
in this excitation-energy region. 
The normalization factors for 
the $0^-$ and $2^-$ components were determined 
so as to reproduce
the known $0^-$ state at $E_x = 9.3~{\rm MeV}$
and the $2^-$ state at $E_x = 4.5~{\rm MeV}$,
respectively, 
while the normalization of the $1^-$ component 
was adjusted to reproduce
the overall magnitude of the bump at 7.5~MeV.
The resulting normalization factors were
0.5, 0.3, and 0.3 for the $0^-$, $1^-$, $2^-$ components,
respectively. 

The resulting calculation, 
shown in Fig.~\ref{fig:accba}, 
reproduces both the cross section
and the tensor analyzing powers reasonably well.
In particular,
the tensor analyzing powers can be close to zero
when a small $0^-$ component is mixed
with a dominant $1^-$ contribution,
because their tensor analyzing powers
have opposite signs and partially cancel each other.
This behaviour explains why the experimental analyzing powers
are similar to those expected for a pure $2^-$ transition,
even though the underlying structure is not dominated by $2^-$.
Together with the absence of a clear peak at $E_x = 7.5~{\rm MeV}$
in the present parity-transfer reaction,
the results suggest that the bump structure
is mainly associated with a $1^-$ excitation
with a smaller admixture of $0^-$ strength. 

\section{Summary}
\label{sec:summary}

In summary,
we have demonstrated that the parity-transfer
$({}^{16}{\rm O},{}^{16}{\rm F}(0^-,{\rm g.s.}))$ reaction 
provides a selective probe
of isovector $0^-$ excitations in nuclei.
The excitation spectrum of $^{12}{\rm B}$
was studied via the
${}^{12}{\rm C}({}^{16}{\rm O},{}^{16}{\rm F}(0^-,{\rm g.s.}))$ reaction
at 247~MeV/u. 
The known $0^-$ state at $E_x = 9.3~{\rm MeV}$
was clearly observed
with an enhanced signal at forward angles, 
confirming the selectivity of the parity-transfer reaction. 

The observed excitation-energy spectrum
and angular distributions are reasonably
described by DWBA calculations 
based on SM transition strengths.
In this description, 
structures observed at
$E_x = 6.6 \pm 0.4$ and $14.8 \pm 0.3~{\rm MeV}$
are suggested to contain significant $0^-$ strength 
and may represent candidates 
for previously unidentified $0^-$ excitations in $^{12}{\rm B}$.

The present results also shed new light on the bump structure
around $E_x = 7.5~{\rm MeV}$ observed
in previous charge-exchange reactions.
Reaction calculations for the $^{12}{\rm C}(d,{}^{2}{\rm He})$ reaction 
indicate that a dominant $1^-$ excitation
with a small admixture of $0^-$ strength
can reproduce both the cross sections and
tensor analyzing powers.
This explains why the structure appears prominently
in conventional charge-exchange reactions but is 
suppressed in the present parity-transfer reaction.

These findings indicate 
the effectiveness of the parity-transfer reaction
as a probe of $0^-$ excitations
and highlight its potential for systematic studies of
spin-isospin modes, including pion-related dynamics,
in atomic nuclei. 

The statistical precision of the present measurement
is lower than in previous charge-exchange experiments
because of the limited intensity of
the ${}^{16}{\rm O}$ beam ($8 \times 10^6~{\rm pps}$), 
which was restricted by radiation-safety regulations at RIBF.
Future experiments with higher-intensity beams would allow
more definitive identification of the $0^-$ candidates in ${}^{12}{\rm B}$, 
and may reveal additional $0^-$ states in other nuclei.

\section*{Acknowledgment}

We thank the accelerator staff
at the RIKEN Nishina Center,
and the CNS, the University of Tokyo, 
for providing us with the excellent beam.
We also would like to thank 
Professor Toshio Suzuki for the helpful discussion. 
This work was supported by JSPS KAKENHI Grants 
No. 17002003, No. 23840053, No. 14J09731, No. 16K17683, and No. 20H01928.


\begin{thebibliography}{9}

\bibitem{Osterfeld1992}
F.~Osterfeld, \RMP{64,491,1992}

\bibitem{Ichimura2006}
M.~Ichimura, H.~Sakai, and T.~Wakasa, \PPNP{56,446,2006}

\bibitem{Fayans1997}
S.~A.~Fayans, E.~E.~Saperstein, and S.~V.~Tolokonnikov, \JPG{3,L51,1977}

\bibitem{Migdal1978}
A.~B.~Migdal, \RMP{50,107,1978}

\bibitem{Meyer-Ter-Vehn1981}
J.~Meyer-Ter-Vehn, \PRP{74,323,1981}

\bibitem{Dozono2008}
M.~Dozono {\it et al.}, \JPSJ{77,014201,2008}

\bibitem{Wakasa2012}
T.~Wakasa {\it et al.}, \PRC{85,064606,2012}

\bibitem{Okamura1995}
H.~Okamura {\it et al.}, \PLB{345,1,1995}

\bibitem{Okamura2002}
H.~Okamura {\it et al.}, \PRC{66,054602,2002}

\bibitem{DeHuu2007}
M.~A.~de~Huu et~al., \PLB{649,35,2007}

\bibitem{Yano2007}
Y.~Yano, \NIMB{261,1009,2007}

\bibitem{Uesaka2012}
T.~Uesaka, S.~Shimoura, and H.~Sakai, \PTEP{2012,03C007,2012}

\bibitem{Michimasa2013}
S.~Michimasa {\it et al.}, \NIMB{317,305,2013}

\bibitem{Dozono2016}
M.~Dozono, T.~Uesaka, S.~Michimasa, M.~Takaki, M.~Kobayashi, M.~Matsushita, 
S.~Ota, H.~Tokieda, and S.~Shimoura, \NIMA{830,233,2016}

\bibitem{Miya2013}
H.~Miya {\it et~al.}, \NIMB{317,701,2013}

\bibitem{HFujita2009}
H.~Fujita {\it et al.}, \PRC{79,024314,2009}

\bibitem{Warburton1992}
E.~K.~Warburton and B.~A.~Brown, \PRC{46,923,1992}

\bibitem{Olsson1993}
N.~Olsson {\it et al.}, \NPA{559,368,1993}

\bibitem{Yang1993}
X.~Yang, {\it et al.}, \PRC{48,1158,1993}

\bibitem{FOLDDWHI}
  J.~Cook and J.~A.~Carr, computer program {\sc fold}, 
  Florida State University (unpublished); 
  based on F.~Petrovich and D.~Stanley, \NPA{275,487,1977};  
  modified as described in J.~Cook, K.~W.~Kemper, P.~V.~Drumm, L.~K.~Fifield, 
  M.~A.~C.~Hotchkis, T.~R.~Ophel, and C.~L.~Woods, 
  \PRC{30,1538,1984}; 
  R.~G.~T.~Zegers, S.~Fracasso, and G.~Col$\grave{\rm o}$, (unpublished).

\bibitem{Franey1985}
M.~A.~Franey and W.~G.~Love, \PRC{31,488,1985}

\bibitem{Furumoto2012}
T.~Furumoto, W.~Horiuchi, M.~Takashina, Y.~Yamamoto, and Y.~Sakuragi, \PRC{85,044607,2012}

\bibitem{Erell1986}
A.~Erell, J.~Alster, J.~Lichtenstadt, M.~A.~Moinester, J.~D.~Bowman, 
M.~D.~Cooper, F.~Irom, H.~S.~Matis, E.~Piasetzky, and U.~Sennhauser, 
\PRC{34,1822,1986}

\bibitem{Inomata1998}
T.~Inomata, {\it et al.}, \PRC{57,3153,1998}

\bibitem{Okamura1999}
H.~Okamura, \PRC{60,064602,1999}

\end{thebibliography}
\end{document}